\documentclass[a4paper,11pt]{article}

\usepackage[utf8]{inputenc}
\usepackage[T1,T2A]{fontenc}
\usepackage{amsmath,amsfonts,amssymb}

 \usepackage{graphicx}
\usepackage{wrapfig}
\usepackage{subfig}
\usepackage{hyperref}
\usepackage[dvipsnames]{xcolor}
\usepackage{dsfont}
\usepackage{amsmath,amssymb}

\begin{document}

\begin{center}
{\LARGE {\bf Warm dark matter from freeze-in at stronger coupling}}
\end{center}

\vspace{1cm}
\begin{center}
{\bf Duarte Feiteira$^{\,1}$, Oleg Lebedev$^{\,1}$, Vinícius Oliveira$^{\,2\,,3\,,4}$}
\end{center}

\begin{center}
  \vspace*{0.25cm}
 $^1$ \it{Department of Physics and Helsinki Institute of Physics,\\
  Gustaf H\"allstr\"omin katu 2a, FI-00014 Helsinki, Finland}\\
  \vspace{0.2cm}
  $^2$ \it{Departamento de Física da Universidade de Aveiro and\\
Center for Research and Development in Mathematics and Applications (CIDMA),\\
Campus de Santiago, 3810-183 Aveiro, Portugal}\\
  \vspace{0.2cm}
  $^3$ \it{Department of Physics, Lund University, SE-223 62 Lund, Sweden}\\
   \vspace{0.2cm}
  $^4$ \it{Laboratório de Instrumentação e Física Experimental de Partículas (LIP),\\
Universidade do Minho, 4710-057 Braga, Portugal}\\
\end{center}

\vspace{2.5cm}

\begin{center} {\bf Abstract} \end{center}
\noindent
We study warm Higgs portal dark matter (DM) in the framework of freeze-in at stronger coupling. This
scenario assumes that the Standard Model thermal bath temperature has always been relatively low, which suppresses  dark matter production.
As a result,  a significant DM-Higgs coupling is allowed, enabling warm dark matter  detection via Higgs decay at colliders.
We find that the Lyman-$\alpha$ bound on the DM mass is particularly strong, excluding masses below 50-100 keV, depending on further details.
The shape of the  DM momentum distribution is highly non-thermal, with low momenta being effectively cut off,  and not captured by the common $\alpha\beta\gamma$-parametrization.

\vspace{1cm}

\newpage

 \tableofcontents
 
 \section{Introduction}
 
 The dark matter (DM) puzzle is one of the outstanding problems of modern physics. The traditional weakly-interacting-massive-particle (WIMP) paradigm finds itself under pressure from the direct DM detection experiments. Thus,
 one is motivated to explore alternative  scenarios involving non-thermal dark matter. A prime example of the latter  is the freeze-in mechanism 
  based on gradual accumulation  of dark matter particles without thermalization \cite{Dodelson:1993je,Hall:2009bx}.  
  This can be  achieved either via small couplings \cite{Hall:2009bx} or 
 via low temperatures \cite{Cosme:2023xpa}, which suppress the reaction rate. 
 
 The traditional high temperature freeze-in framework suffers from the background of gravitationally produced particles. Once produced, feebly interacting stable particles remain and accumulate. 
 Both classical and quantum gravitational effects  during and immediately after inflation provide an efficient particle production mechanism,  
  which often results in dark matter over-abundance \cite{Lebedev:2022cic}. To give an example,  
 gravity induces   Planck-suppressed couplings between the inflaton field   $\phi$ and dark matter, e.g. \cite{Lebedev:2022ljz,Koutroulis:2023fgp}
\begin{equation}
{1\over M_{\rm Pl}^2}\, \phi^4 S^2 ~,~ {1\over M_{\rm Pl}}\, \phi^2 \bar \chi \chi ~,...
\end{equation} 
Here $S$ and $\chi$ are a scalar and a fermion,  respectively. During the inflaton oscillation epoch, such couplings lead to efficient production of these particles that generally overtakes the traditional freeze-in mechanism.
A similar effect is created by the inflationary de Sitter fluctuations \cite{Feiteira:2025rpe,Costa:2026wxm}, which eventually turn into particles, although this production channel  is  much weaker for fermions \cite{Feiteira:2025phi}.

The problem can be   solved in scenarios with a low reheating temperature $T_R$  \cite{Lebedev:2022cic}. The energy density of the produced particles gets diluted by the Universe expansion driven by a non-relativistic inflaton. If the latter decays at late times,
the dark relic abundance gets suppressed. 
The corresponding reheating temperature can even be below the dark matter mass \cite{Cosme:2023xpa}.
 This possibility has recently attracted considerable  attention   \cite{Bernal:2026clv}-\cite{Henrich:2024rux}    due to its interesting observational prospects in collider or direct detection experiments.
 The DM production rate  is Boltzmann--suppressed allowing for a significant coupling to the SM fields and, thus,  possible detection of freeze-in dark matter \cite{Cosme:2023xpa}.

 In our current work, we focus on another  regime of the low temperature freeze-in. Dark matter itself can be very light, but the SM particles responsible for its production could have masses above the bath temperature.
 We find that this possibility is viable and can account for warm dark matter in the 100 keV mass range. Its coupling to the Higgs field can be significant and lead to invisible Higgs decay at the LHC and future colliders.

 \subsection{Higgs portal dark matter}
 
 In what follows, we study light dark matter in the Higgs portal framework \cite{Patt:2006fw,Lebedev:2021xey} with a low reheating temperature. Let us define the model.

 The simplest dark matter model  is a real scalar extension of the SM, where the DM stability is enforced by a $Z_2$ symmetry \cite{Silveira:1985rk}.  
  The interaction between the SM sector  and dark matter $S$ is given by 
 \begin{eqnarray}
{\cal L}_{hs} &=& {\lambda_{hs} \over 2}\; H^\dagger H SS  \;. 
\label{Higgs-portal1}
\end{eqnarray}
 In the unitary gauge, the Higgs doublet is replaced by $(h+v)/\sqrt{2}$, where $v$ is the Higgs VEV and $h$ is the physical Higgs boson.
 Its interactions relevant for our purposes are 
  \begin{eqnarray}
 && {\cal L}_{hss}= {\lambda_{hs} v \over 2}\; hSS \;, \\
 && {\cal L}_{SM}= {m_f\over v}\; h \bar f f  \;.
\end{eqnarray}
The resulting DM production mechanism is of the freeze-in type as long as the SM bath temperature is sufficiently low, which can be probed via invisible Higgs decay  \cite{Lebedev:2024mbj,Arcadi:2024wwg}.
We note that the physical scalar mass depends on the bare mass term $\mu^2 S^2$ and can, in principle, be arbitrarily light (see, for example, Sec. 3.2 of \cite{Lebedev:2011aq}). In order to obtain a very light scalar, one requires a special relation among the (renormalized) quantities in the scalar potential,
\begin{equation}
{1\over 2} \lambda_{hs} v^2 \simeq - \mu^2 \;.
\end{equation}
In this work, we do not address the origin of this relation, which can only be done within a more fundamental framework. Instead, we explore the available parameter space consistent with all the constraints.

We are interested in very light dark matter with mass below 1 MeV.
It can efficiently be produced via the freeze-in mechanism even at low temperatures, $T \lesssim 100\,$MeV.
Annihilation of light SM particles mediated by the Higgs boson leads to the observed DM relic abundance as long as the coupling $\lambda_{hs}$ is not too small.
The production rate is suppressed by the light fermion Yukawa couplings as well as by the Higgs mass such that 
DM never thermalizes and 
the model belongs to the freeze-in category.

\begin{figure}[h!]
    \centering
    \includegraphics[width=0.77\textwidth]{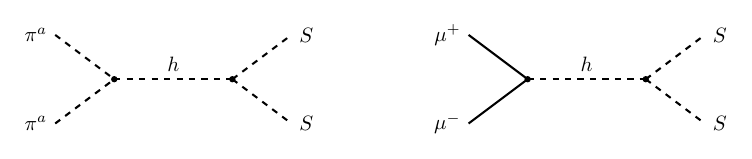}
    \caption{Dark matter production via pion and muon annihilation.   }
    \label{diagram}
\end{figure}

 The leading DM production channels are shown in Fig.\,\ref{diagram},
  which dominate for $T \lesssim 100\,$MeV. 
  This upper limit is dictated by the proximity of the QCD phase transition, close to which the perturbative approach breaks down.
  At lower temperatures, one may treat the hadrons in the dilute gas approximation \cite{HotQCD:2012fhj}, with the pions and muons playing the main role.
  As their abundance gets further Boltzmann-suppressed, the electron annihilation channel starts to dominate at 
  $T\sim 10\,$MeV.

   The framework of freeze-in at stronger coupling assumes that the SM thermal bath temperature has never been high. This is consistent with the observational bounds which only require the reheating temperature to be above 4 MeV \cite{Hannestad:2004px}.
   In the next section, we discuss classes of postinflationary scenarios, in which the maximal temperature of the SM sector does not significantly exceed the reheating temperature. Such models do not require non-standard cosmological evolution and
    only relax the traditional assumption of how the SM sector is produced after inflation.

 \section{Temperature evolution in the Early Universe}
 
 The exact mechanism of the Standard Model field production in the Early Universe remains unknown. A simple  possibility is that the observable sector was produced via decay of 
 some primordial field, for example, an inflaton      \cite{Kolb:1990vq}     or a generic scalar \cite{Cosme:2024ndc}. The resulting SM energy density  and  temperature depend on properties of this field.
 It is thus  reasonable to parametrize the uncertainty 
in terms of a few  scaling parameters, which vary from model to model.

 Consider a general case of the SM sector production from decay of a field $\chi$, which does not necessarily dominate the energy density of the Universe \cite{Cosme:2024ndc}. 
  The SM quanta are normally  produced in the relativistic regime, so we can assume that the SM energy density  $\rho$ scales as radiation. 
 The energy density of the $\chi$ field is  denoted by $\rho_\chi$, whose    scaling  with the scale factor is   $a^{-n}$, for  some positive $n$. 
 The Hubble rate $H$ scales as  $a^{-m}$, with $m>0$.
  The values of $n$ and $m$ are, in general, unrelated and depend on further details. 
  
  Denoting the decay width of $\chi$ by 
 $\Gamma_\chi$, we arrive at the following system of equations:
\begin{eqnarray}
&& \dot \rho + 4 H\rho= \Gamma_\chi \rho_\chi \;,  \nonumber \\
&& H= H_0 /a^m \;, \nonumber \\
&& \rho_\chi = \rho_\chi^0 / a^n \;. \label{system}
\end{eqnarray}
The label  $0$ refers to the initial moment  $a_0=1$  which marks the end of inflation and $\rho (1)=0$. 
 Here we assume  that the SM sector does not contribute significantly to the energy balance of the Universe $before$ reheating
 and that $n,m$ are constant in the regime of interest.

The solution for $\rho(a)$ is easily found using  $\dot \rho + 4 H\rho = {1\over a^4} \, { d\over dt} (a^4 \rho)$ and $dt = da/(aH)$.  Then,
\begin{equation}
a^4 \rho =  \Gamma_\chi \rho_\chi^0\; \int {da \over aH} \,a^{4-n} \;,
\end{equation}
and the boundary condition $\rho (1)=0$ requires \cite{Cosme:2024ndc} 
\begin{equation}
  \rho(a) =  {\Gamma_\chi \rho_\chi^0 \over  (4-n+m)H_0}\; \left[    {1\over a^{n-m}} -{1\over a^4}  \right] ~~\rightarrow ~~{\Gamma_\chi \rho_\chi^0 \over  (4-n+m)H_0 } \; {1\over a^{n-m}} 
  \label{rhoSM}
\end{equation}
at $a\gg 1$, given that   $n-m <4$ for all cases of interest. The  approximation becomes reasonable already at $a\sim  {\cal O}(1)$. The  scaling parameter $n-m$
can have either sign or be zero, depending on properties of $\chi$. In particular, the SM energy density would $increase$ over time if $n-m<0$.

In the simplest case, the inflaton  $\phi$ can directly decay into the SM states and the role of the mother particle $\chi$ is played by the inflaton itself  \cite{Kolb:1990vq}.
Since it dominates the energy balance, $m=n/2$ and the SM energy density scales as $a^{-m}$, with $m $ being $3/2$ or $2$ for the non-relativistic and relativistic inflaton, respectively.
This scaling is maintained until reheating, after which the usual result $\rho \propto a^{-4}$ applies. 
 
 The Standard Model  fields start being produced immediately after inflation. Due to the large couplings, they thermalize rather quickly and, soon after their production, the SM sector  can be assigned the temperature
\begin{equation}
T_{\rm SM}\simeq       \left({30\over g_* \pi^2}\right)^{1/4}   \rho^{1/4}  ~   \propto     ~  { a^{-{n-m\over 4}}}\;,
\end{equation} 
where $g_* $ is the number of the SM degrees of freedom. In the simplest case, this temperature decreases as $a^{-n/8}$ until reheating, when the observable sector takes over the energy balance. Therefore, if the SM fields are produced via $\phi \rightarrow {\rm SM}$, the maximal  temperature is far above the 
reheating temperature.

\begin{figure}[h!]
    \centering
    \includegraphics[width=0.59\textwidth]{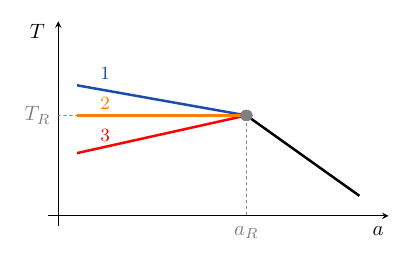}
    \caption{  Temperature   of the Standard Model sector in the postinflationary epoch (on a logarithmic scale), with reheating
    at   $a\sim a_R$.
    The different scenarios represent: (1) direct SM particle production through inflaton decay, $n-m>0$, 
    (2) SM sector production via a cascade decay, $n-m=0$, (3) SM sector production via a longer cascade decay, $n-m<0$.
    }
    \label{Tmax}
\end{figure}

In a more general situation, $n-m$ differs from $n/2$ and can take on negative or zero values. 
For example, the inflaton decay into the SM sector can be blocked by symmetry requirements, while $\phi \rightarrow \chi \chi$ would be allowed. The observable sector is then produced via decay of the dark state $\chi$,
\begin{equation}
\phi \rightarrow \chi \chi \rightarrow {\rm SM}\;,
\end{equation}
 or a more complicated decay chain.
This happens, for example, when the inflaton is charged under an (approximate) lepton number symmetry such that it decays primarily into the right-handed neutrinos via the coupling $\phi \, \nu_R \nu_R$.  
If these are very weakly coupled and do not thermalize, they can play the role of $\chi$ and produce  the SM states through $\nu_R \rightarrow \ell H$. 
According to the argument above, the energy density of $\nu_R$ scales exactly as the Hubble rate does, $a^{-m}$, which implies
\begin{equation}
n-m =0\;,
\end{equation}
and the SM temperature remains constant until reheating. As a result,  the maximal temperature is close to the reheating temperature, $T_{\rm max} \simeq T_R$.

The value of $n$ decreases as the decay chain gets longer, $\phi \rightarrow \chi_1 \chi_1 \rightarrow \chi_2 \chi_2 \chi_2 \chi_2 \rightarrow ...$, while $m$ is constant as long as the energy density is dominated by the mother particle. Thus, $n-m<0$ if the decay chain contains more
than one intermediate particle.  In this case, the SM temperature increases until reheating, after which it decreases as $a^{-1}$.  
In the transition region, the energy balance changes and the approximation of the inflaton domination becomes inadequate. However, the transition to reheating can happen 
quite  fast, depending on the decay rates of $\chi_i$, such that $T_{\rm max}$ would remain close to $T_R$.

The different possibilities for the SM temperature evolution are illustrated in Fig.\,\ref{Tmax}.  Option (1) corresponds to the traditional inflaton decay into the SM states, while options (2) and (3) reflect more general 
production mechanisms involving decay of intermediate particles. The blob at $a\sim a_R$ reflects the uncertainty in the transition region. Here, ``reheating'' is defined as the  period after which the energy density gets 
dominated by the SM thermal bath.

In cases (2) and (3), the temperature of the SM bath  may have never exceeded the MeV scale. Indeed, the lower bound on the  reheating temperature is about 4 MeV.
  Such a low temperature  {\it does not require} non-standard cosmology, it only necessitates small couplings of the primordial fields to the SM particles.
  This class of models is the focus of our present work.
  We note that the $n-m >0$ case corresponds to the traditional situation with $T_{\rm max} \gg T_R$ and DM production not being Boltzmann-suppressed.
 
We conclude that, since the SM particle production mechanism remains unknown,  one ought to account for distinct  possibilities for the temperature evolution in the Early Universe. In the context of freeze-in production of dark matter, they lead 
to vastly different scenarios. For example,  when $T_{\rm max} \simeq T_R$, production of dark matter with mass above $T_R$ is dominated by the temperature range $T\simeq T_R$. 
This distinguishes the 
 low $T_R$ freeze-in models from their more traditional counterparts, leading to 
   distinct  phenomenology.

\section{Freeze-in at very low temperatures}
 
 Since the observational bound on the reheating temperature is about 4 MeV \cite{Hannestad:2004px}, it is 
  possible that the temperature of the Standard Model thermal bath has never been high. In fact, it could be lower than the dark matter mass or the mass of the SM mother particle  responsible 
 for DM production. 
  In these cases,  the production rates are Boltzmann-suppressed. As a result, low-T freeze-in is compatible with 
  significant DM-SM couplings,  without dark matter thermalization \cite{Cosme:2023xpa}.

The low $T_R$ framework allows one to dilute any pre-existing  DM abundance produced at the early stages of the Universe evolution \cite{Cosme:2023xpa}
, e.g. during preheating. Indeed, as  long as the Universe energy density is dominated by the non-relativistic inflaton, the DM energy fraction decreases as $1/a$
such that it becomes negligible at late times. Thus, if reheating occurs at sufficiently low $T$, the initial DM abundance can be neglected.

 In this work, we are interested in warm dark matter, which is intrinsically very light. Thus,  Boltzmann suppression appears  due to the bath temperature being lower than the SM particle masses responsible for DM production.
In the Higgs portal models,    such  mother particles are pions and muons  in much of the parameter space of interest \cite{Lebedev:2024mbj}.

 We focus on temperatures below 100 MeV such that the pion approximation  of the SM hadron gas becomes adequate. 
 This is due to the pions being relatively dilute and the momentum transfer in the collisions being small, enabling the EFT approach.
 At higher temperatures, close to the QCD phase transition scale,  the SM bath cannot be treated 
 in the weakly coupled gas approximation \cite{HotQCD:2012fhj}. However, at yet higher temperatures, the heavy quark and lepton production channels dominate, making the perturbative calculation reliable again \cite{Lebedev:2024mbj}.
 
 In the above temperature range, the particles with the largest couplings to the Higgs are pions and muons. They give comparable contributions to the DM production.
 At very low temperatures, $T \sim 10\,$MeV, both of them become irrelevant due to the strong Boltzmann suppression and the production is dominated by the electron channel.

 In the next subsection, we consider the kinematics of light dark matter production in a low temperature thermal bath. This analysis is instructive for understanding main features of the  momentum distribution function.

 \subsection{Production kinematics}
 \label{kinematics}
 
 Consider the light dark matter production via pion annihilation in the regime $m_s\ll T\ll m_\pi$. The pions are mostly non-relativistic, so the DM energy is naturally close to $m_\pi$. Hence, the bulk of the produced DM particles would have a momentum distribution
 centered around $m_\pi$. On the other hand, the pions at the Boltzmann tail produce highly energetic as well as low energy particles.  At first glance, it appears that the DM momenta should be cutoff at $|{\bf p}|=m_\pi$
 from below, yet certain kinematic configurations lead to production of DM particles with energy below the pion mass.

Let us neglect the DM mass ($m_s\rightarrow 0$) for now and consider the reaction
$$    k_1 +k_2 \rightarrow \tilde p + p     \;,  $$
where $k_i$ are the pion 4-momenta, while $p,\tilde p$ are the DM 4-momenta in the thermal bath rest (``lab'') frame. Suppose we are interested in large $|{\bf p}|\gg m_\pi$. The initial state pion(s) must  then be highly relativistic.
The reaction rate contains the usual Boltzmann factors for the initial state,
\begin{equation}
e^{-E_1/T} \,e^{-E_2/T}\;,
\end{equation}
where $E_i$ are the temporal components of $k_i$.
For the reaction to be most efficient, the Boltzmann suppression must be minimized,
$$ E_1 + E_2 \rightarrow {\rm min} \;,$$ 
subject to the constraint that $ E_1 + E_2 \geq p_0$. This requires minimizing $\tilde p_0$, while in the center-of-mass (CM) frame, the DM particle energy is bounded by $m_\pi$ from below.
Clearly, the necessary particle energy reduction can only be achieved via a strong boost in the direction  of motion of the tilded particle. 

\begin{figure}[h!]
    \centering
    \includegraphics[width=0.59\textwidth]{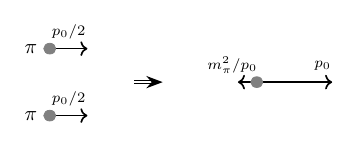}
    \caption{  Production of one hard  and one soft  DM particle via a relativistic pion collision.  Corrections of order $m_\pi^2/p_0$ to the pion momenta are neglected.     }
    \label{kin}
\end{figure}

Let us formalize this argument. 
The minimal energy DM configuration in the CM frame is 
\begin{equation}
p'=\left(
 \begin{matrix}
  m_\pi \\
 m_\pi
 \end{matrix}
 \right)~~,~~ 
 \tilde p' = \left(
 \begin{matrix}
  m_\pi \\
 -m_\pi
 \end{matrix}
 \right)\;,
\end{equation}
where the spacial component of the 4-momentum refers to the $z$-direction.
The boost to the lab frame is accomplished by
\begin{equation}
\Lambda = 
\left(
 \begin{matrix}
\cosh  \eta & \sinh \eta \\
 \sinh \eta  & \cosh \eta
 \end{matrix}
 \right) ~~,~~ p= \Lambda \left(
 \begin{matrix}
  m_\pi \\
 m_\pi
 \end{matrix}
 \right) =  e^\eta\;\left(
 \begin{matrix}
   m_\pi \\
  m_\pi
 \end{matrix}
 \right) = \left(
 \begin{matrix}
  p_0\\
  p_0
 \end{matrix}
 \right)~,
\end{equation}
where $\eta$ is the rapidity.
This fixes
\begin{equation}
e^\eta = p_0/m_\pi ~~,~~ \tilde p =  m_\pi/p_0 \;\left(
 \begin{matrix}
   m_\pi \\
  -m_\pi
 \end{matrix}
 \right) \rightarrow 0\;
\end{equation}
at large $p_0$.
Thus, we get
$ E_1 +E_2 = p_0 + m_\pi^2/p_0$,
which tends to $p_0$ at large energies.\footnote{If the CM energy is not minimized by the DM configuration, the correction behaves as $E^2/p_0$ instead of $m_\pi^2/p_0$, with $E> m_\pi$.}
 This configuration  minimizes the Boltzmann suppression for a given $p_0 \gg m_\pi$,
 \begin{equation}
e^{-E_1/T} \,e^{-E_2/T}= \exp \left(    -{p_0\over T} - {m_\pi^2 \over {p_0T}}      \right)\;.
\label{Bf}
\end{equation}
 It is this combination of  the momenta that appears in the asymptotic limits of the Boltzmann collision term. The corresponding kinematic configuration is shown in Fig.\,\ref{kin}. The pion momenta are found by boosting their CM values by $\Lambda$, which yields
 approximately ${1\over 2}( p_0   , p_0  )^T$, up to $m_\pi^2/p_0$ corrections.
 
This exercise also shows that DM with arbitrarily small momenta can be produced, although the production rate is strongly Bolztmann-suppressed.
By relabelling 
$ p_0     \leftrightarrow m_\pi^2/p_0 $, one finds that the corresponding Boltzmann factor is again given by (\ref{Bf}) with $p_0 \ll m_\pi$.

\subsection{Boltzmann equation and the collision integral}

The critical quantity for our analysis is 
the dark matter momentum distribution function $f(|{\bf p}|,t)$. It  satisfies the Boltzmann equation \cite{Kolb:1990vq}
\begin{equation}
{\partial f\over \partial t} - H |{\bf p}| {\partial f\over \partial |{\bf p}|} = {1\over p_0}\;{\cal C}(p,t) \;,
\end{equation}
where $H$ is the Hubble rate and ${\cal C}(p,t)$ is the collision term describing particle production.
The left hand side of this equation is equivalent to
$   {d\over dt} f \left( |{\bf p}(0)|   \, {a_0\over a(t)} , t  \right) , $
so introducing the physical momentum
$ {\bf p}(t) \equiv    {a_0\over a(t)} \,  {\bf p}(0)    \, ,$
one  can integrate the equation with the result
\begin{equation}
f( |{\bf p}(t)|, t) =\int dt' \, {{\cal C}(p',t') \over p_0'} ~~{\rm with}~~ {\bf p}' = {{\bf p}(t) \, {a(t) \over a(t')}}\;.
\end{equation}
Particle production is controlled by the collision integral,
 \begin{equation}
 {\cal C} (p)=   \,{1\over (2\pi)^9}\, \int {d^3 {\bf k}_1 \over 2 E_{1}} {d^3 {\bf k}_2 \over 2 E_{2}} {d^3 {\bf \tilde p} \over 2 E_{\tilde p}} \; (2\pi)^4 \delta^{(4)} \left(   k_1+k_2-\tilde p - p    \right) \, \vert {\cal M} \vert^2 \, f(k_1) f(k_2)\;,
 \end{equation}
 with the Maxwell-Boltzmann statistics  
 $    f(k_1) f(k_2) = e^{-{(E_1+E_2) / T }}    $.
  Here we follow the definition of ${\cal C} (p)$ from \cite{DEramo:2020gpr}.
 For the leading Higgs portal reactions, the scattering matrix element is a function of the CM energy,
 $  \vert {\cal M} \vert^2 = \vert {\cal M} (s) \vert^2 , ~ s\equiv  (k_1+k_2)^2 \;.   $
 The integral above is { Lorentz-invariant} apart from the thermal distribution factor $f(k_1) f(k_2) $.

The collision term has been computed, for example,  in \cite{DEramo:2020gpr}.  
To understand the shape of the DM momentum distribution, let us consider the asymptotic regimes of large and small momenta.
Focussing on the pion contribution at $T\ll m_\pi$, we can set 
 $  \vert {\cal M} (s)\vert^2 \simeq  \vert {\cal M} (4m_\pi^2) \vert^2$, since the annihilation proceeds primarily via the $s$-wave.\footnote{Relaxing this assumption does not affect the collision integral in any significant way since 
 its behavior is controlled primarily by the exponential factor. The   $s$-dependence of the amplitude is given in \cite{Winkler:2018qyg}, although it cannot be extrapolated to energies far above the pion mass. }
  This applies also to the asymptotic regimes considered in Sec.\,\ref{kinematics}:
 even though $p_0 \gg m_\pi$ or $p_0 \ll m_\pi$, the CM energy is still close to $2 m_\pi$.
 Denoting $p \equiv |{\bf p}|$ and setting $m_s\simeq 0$, we have 
 \begin{equation}
{\cal C} (p) \simeq  c\; {T e^{-p/T}\over p}  \; \int_{4 m_\pi^2}^\infty ds  \: \sqrt{1-4m_\pi^2/s} \; e^{-{s\over 4pT}}  \;,
\end{equation}
where $c$ is a constant proportional to $\vert {\cal M} (4m_\pi^2) \vert^2$.
The integral is easily evaluated using  the integration variable $x=(s-4m_\pi^2) /4pT$ and  one  finds the following asymptotic limits:
\begin{eqnarray}
&&  p\ll m_\pi^2/T:~~{\cal C} (p)  \propto  {p^{1/2}T^{5/2}\over m_\pi} \; \exp \left(   -{p\over T} -{m_\pi^2\over pT} \right)  \;, \nonumber \\ 
&& p\gg m_\pi^2/T: ~~ {\cal C} (p)  \propto  T^2 \; \exp \left(   -{p\over T} -{m_\pi^2\over pT} \right) \;.
\end{eqnarray}
These results are consistent with the  considerations of Sec.\,\ref{kinematics}.  In particular, the form of the exponential is dictated by the collision kinematics.
Similar considerations   apply to the muon annihilation channel, up to inconsequential modifications.

The consequent momentum distribution function depends on the temperature evolution in the Early Universe. In what follows, we consider two possibilities which motivate the concept of freeze-in at stronger coupling:  one, in which the temperature peaks at 
$T\simeq T_R$, and another with a flat temperature profile before reheating.

 \subsection{Models with a temperature peak at $ T_R$}
 
 Consider a class of models 
  with $n-m<0$ in Eq.\,\ref{system}, i.e. curve (3) in Fig.\,\ref{Tmax}. In many aspects, the physics of dark matter production in such scenarios is captured by the {\it instant reheating approximation}.
  Indeed, the DM production before reheating is suppressed by both  the Boltzmann and dilution  factors. Hence, one may neglect it and simply approximate the system by the effective temperature $T=0$ before reheating and 
  $T\propto T_R /a$ after reheating. The quality of this approximation depends on the value of $n-m$ as well as further details. Nevertheless, we will adopt it in what follows, keeping in mind the associated uncertainty.
  
 We note that our results can readily be extended  to a more general situation with $T_{\rm max} > T_R$ as long as $T_{\max} $ is low enough to ensure the Boltzmann suppression of DM production \cite{Cosme:2024ndc}. 
  
  The main features of the distribution function can be understood analytically. 
   Introducing the comoving momentum
  \begin{equation}
  q\equiv {p\over T} \;,
  \end{equation}
  we have 
  \begin{equation}
f(q) =   {1\over q} \; \int_T^{T_R} {dT\over T^2}\, {{\cal C} (q, T)   \over H}  \;,
\end{equation}
  taking  the number of the SM degrees of freedom approximately constant. The main difference from the standard freeze-in calculations, which assume a very high reheating temperature, is that the upper limit of integration is 
  rather low instead of  $\infty$. This affects the functional dependence of $f(q)$. Typically it takes the form $e^{-q}/\sqrt{q}$ in high $T$ models, whereas in our case it becomes quite different.
  
  Focussing again on the pion contribution, the integral can easily be computed in the asymptotic regions:
  \begin{eqnarray}
  && q\ll m_\pi^2/T_R^2~~:~~  f(q) \propto   \sqrt{q}   \; \exp\left(    -{m_\pi^2 \over q T_R^2}   \right) \;,\nonumber \\
  && q\gg m_\pi^2/T_R^2 ~~:~~  f(q) \propto  {1\over \sqrt{q} }\; \exp\left(   -q  \right)\;.
  \label{f(q)}
  \end{eqnarray}
  The large momentum behavior remains the same as in the high $T$ models, wheres at low momenta, it changes completely. The IR modes get effectively cut-off.
   The reason is clear:  at the time of production, the physical momenta are  close to $m_\pi$, while at later times they red-shift
  by the factor $T/T_R$. 
  The particles simply do not have enough time to red-shift to lower values, resulting in strong suppression of the IR modes. The typical comoving momentum is $q\sim m_\pi/T_R $ and the lower values
  like $q\sim 1$ would be exponentially suppressed.\footnote{The modes with $q\sim m_\pi/T_R \ll (m_\pi/T_R)^2 $ appear to be suppressed according to the asymptotic expression, however, the   value
  of $f(q)$ depends on the overall normalization, e.g. the Higgs portal coupling, and thus can be adjusted at will.}  In high-$T$ models, the red-shifting is effective over a much longer period and the above feature is thus absent.
 
\begin{figure}[h!]
    \centering
    \includegraphics[width=0.55\textwidth]{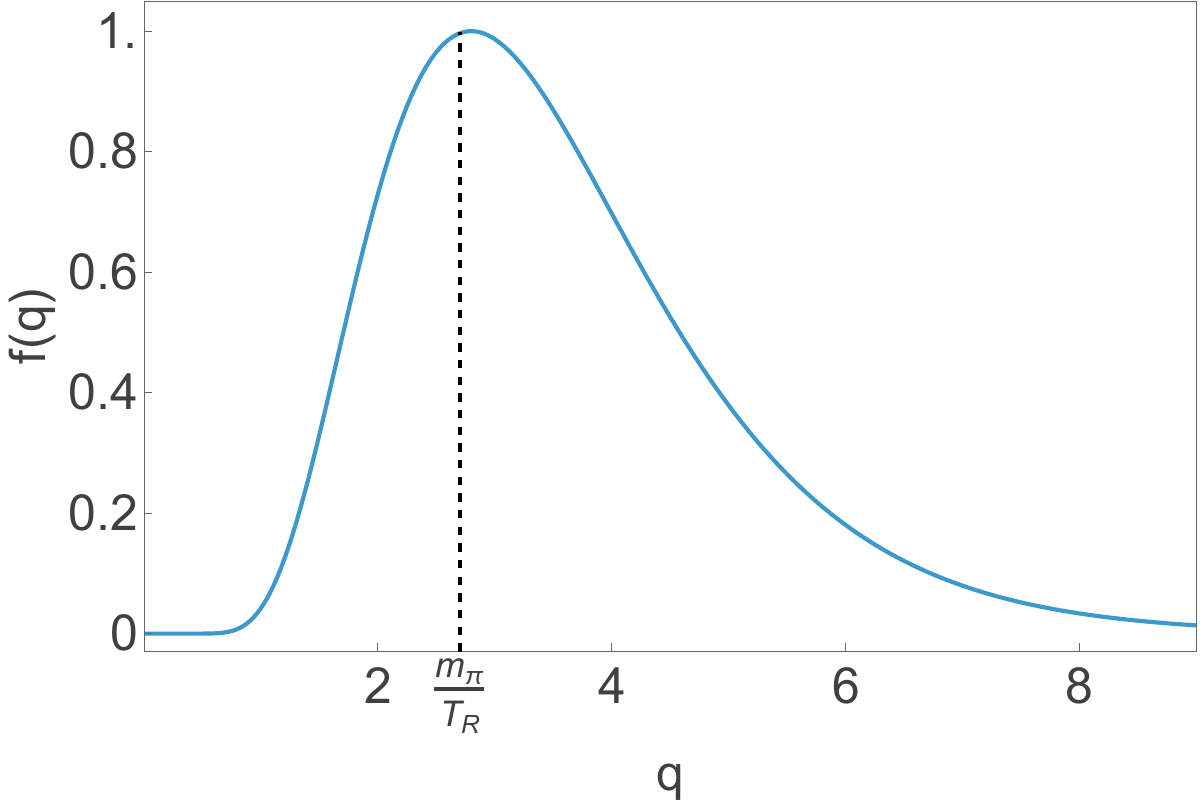}
    \caption{Comoving momentum distribution function of DM produced via pion annihilation, with $T_R=50\;$MeV and assuming instant-like reheating.   
    For convenience, $f(q)$  is normalized to 1 at the maximum.}
    \label{f}
\end{figure}

 Our numerical integration result is presented in Fig.\,\ref{f}.  It is clearly seen that the low momenta are exponentially suppressed, while the typical value is $q\sim m_\pi/T_R$. 
 These features make the distribution highly non-thermal and  
 shock-wave-like,  distinguishing  it from $f(q)$ the high-$T$ freeze-in models \cite{Ballesteros:2020adh}-\cite{DEramo:2025jsb}. 
 
 In the muon annihilation channel, we find a  similar distribution function due to the proximity of the pion and muon masses. The electron counterpart is quite different, yet the electron contribution is suppressed and does not play any significant role.

 \subsubsection{On the $\alpha\beta\gamma$ parametrization}
 
  It is common to parametrize the momentum distribution function as \cite{Bae:2017dpt}
  \begin{equation}
  f(q)\Big\vert_{{\rm high}\, T_R} \propto q^\alpha \; \exp(\beta\, q^\gamma) \;,
  \end{equation}
  with constant $\alpha,\beta,\gamma$. Although the high $T_R$ freeze-in models are accommodated by this Ansatz very well,  low $T_R$ scenarios 
do   not follow the same pattern. Indeed, the asymptotic form  (\ref{f(q)}) shows that the coefficients $\alpha,\beta,\gamma$ become momentum-dependent,
$$    \alpha,\,\beta,\,\gamma~ \rightarrow~ \alpha(q),\,\beta(q),\,\gamma (q)  \,.  $$
 For example,
$\gamma$ flips from 1 at large momenta to -1 at low momenta. Hence, the functional form of $f(q)$ is quite different and the      $\alpha,\beta,\gamma$ parametrization
is not particularly useful. Nor does the above distribution fit the quasi-thermal form \cite{DEramo:2025jsb}.

We find that this feature is generic in the  low $T_R$ framework  and not restricted to the Higgs portal models. For example, DM can be produced via decay of a mother particle at low $T$.
In this case,  the kinematic considerations of   Sec.\,\ref{kinematics}
apply with some modifications, while the main features remain. In particular, the low momentum states are produced by a highly boosted mother particle, which leads to strong Boltzmann-suppression.
As a result, $f(q)$ at small $q$ exhibits a similar behavior to that observed in our case. Furthermore, in reality, dark matter can be produced by multiple mother particles, which can lead to an even more complicated 
non-$\alpha\beta\gamma$
shape of $f(q)$, depending on their  relative contributions.

  We emphasize that these conclusions are reached in the framework of {\it Standard Cosmology} after reheating.  In particular,  the expansion history has not been modified. We only assume that the reheating temperature 
  is low due to small couplings and that the SM states have been produced via decay of a spectator field. 
  
    Strong suppression of the low-momentum modes leads to strict bounds on warm dark matter from structure formation \cite{Viel:2005qj}, which we quantify in the next subsection.

    \subsubsection{Dark matter abundance}

   The dark matter abundance is computed via the   Boltzmann equation for the number density $n$,
    \begin{eqnarray}
\dot n +3Hn & = & 2 \,\Gamma ({\rm SM \rightarrow DM}) - 2\,\Gamma ({\rm DM \rightarrow SM})  \;,
 \end{eqnarray}
where $\Gamma$ denotes the reaction rate per unit volume and the factor of 2 appears due to 2 DM quanta being involved in each reaction.
 In the freeze-in framework, the produced DM density is far below its thermal equilibrium value such that  DM annihilation can be neglected, $\Gamma ({\rm DM \rightarrow SM}) \simeq 0$. 
 For the Maxwell-Boltzmann distribution function, the production reaction rate takes the form
\begin{equation}
\Gamma(\bar f f \rightarrow SS) = \langle  \sigma (\bar f f \rightarrow SS) v_r \rangle \, (n_f^{\rm eq})^2 \, , 
\end{equation}
where $\langle ... \rangle$ stands for the thermal average, $\sigma$ is the reaction cross section which includes the symmetry factors for {\it the initial and final}  states, $v_r$ is the relative velocity of the initial state particles and
$n_f^{\rm eq}$ is the thermal equilibrium density of $f$.

   At low temperatures, light dark matter can be produced via annihilation of
   \begin{itemize}
  \item{muons}
  \item{pions}
  \item{electrons}
  \item{photons} 
   \end{itemize}
   Numerically, the dominant  contributions are provided by the muons and the pions. Although the muon annihilation amplitude suffers from mild  velocity suppression, the muon density is higher, which makes the reaction rate comparable to or higher than that 
   for the pions.
   The electron contribution is strongly suppressed by the electron mass and even though the electron density is high, the corresponding reaction rate is  small. It 
   starts to dominate only at very low temperatures, $T\sim 10\,$MeV, when the heavier particles become very dilute. On the other hand, we find that the photon channel is always negligible \cite{Lebedev:2024mbj}.

   The production cross section calculation  is based on the following scattering amplitudes.
   The pion annihilation occurs  primarily  at low momentum transfer, i.e. $s\simeq 4m_\pi^2 $. This regime is captured by the chiral perturbation theory and the low energy theorems \cite{Vainshtein:1980ea,Voloshin:1986hp,Dawson:1989yh},  and the amplitude is extracted from    
    $\vert {\cal M}_{h\rightarrow {\pi^i \pi^i}} \vert = \vert {\cal M}_{{\pi^i \pi^i}\rightarrow h} \vert$ at $s\simeq 4m_\pi^2 $, using the results of \cite{Winkler:2018qyg}.
    For a single pion type,  this yields 
   \begin{equation}
   \vert {\cal M}_{{\pi^i \pi^i}\rightarrow SS} \vert^2 \simeq   {4 \, \lambda_{hs}^2 m_\pi^4 \over m_h^4}  \;,
   \end{equation}
   in the {\it standard QFT convention}, for $m_s \sim 0$. When this matrix element enters the thermal rate calculation, it is to be multiplied by the phase space symmetry and multiplicity factors.
   Due to the  identical pions in the initial state and identical scalars in the final state, the symmetry factor is $1/2 \times 1/2$, while the pion multiplicity factor is 3. 
   
   The standard QFT muon annihilation amplitude squared, averaged over the initial state spins, is given by 
   \begin{equation}
    \left|{\overline {\cal M}}_{\mu^+ \mu^- \rightarrow SS}\right|^2  =    {\lambda_{hs}^2 m_\mu^2  (s-4m_\mu^2) \over 2 m_h^4}\;,
   \end{equation}
   for $m_s \sim 0$.
      In the thermal reaction rate, this matrix element gets multiplied by the spin multiplicity factor $2^2$ as well as the phase space symmetry factor 1/2 due to identical scalars in the final state.
   
   The Boltzmann equation is solved numerically and the DM abundance $Y$ is computed via
   \begin{equation}
 Y= {n\over s_{\rm SM}} \;,
 \end{equation}
 where $s_{\rm SM}$ is the SM entropy density, 
  \begin{equation}
s_{\rm SM}= {2\pi^2 \over 45} g_s T^3~,~ H = \sqrt{g_* \pi^2 \over 90 } \, {T^2 \over M_{\rm Pl}} \;.
\end{equation}
Here $g_s$ is the SM d.o.f. number contributing to the entropy and $g_*$ is that  contributing to the energy density. Due to entropy conservation, $Y$ remains constant after DM production completes, which in practice means  
shortly after $T \sim T_R$. The observed value is 
\begin{equation}
Y_\infty = 4.4 \times 10^{-10}\; {{\rm GeV}\over m_s}\;.
 \end{equation}
 In the regime $m_s \ll m_{\pi,\mu}$, the reaction rates and the resulting $n$ are independent of the DM mass. Since $\Gamma, \,n \propto \lambda_{hs}^2$, we therefore find the scaling
 \begin{equation}
 \lambda_{hs} \propto {1\over \sqrt{m_s}} \;.
 \end{equation}
 
   \begin{figure}[h!]
    \centering
    \includegraphics[width=0.5\textwidth]{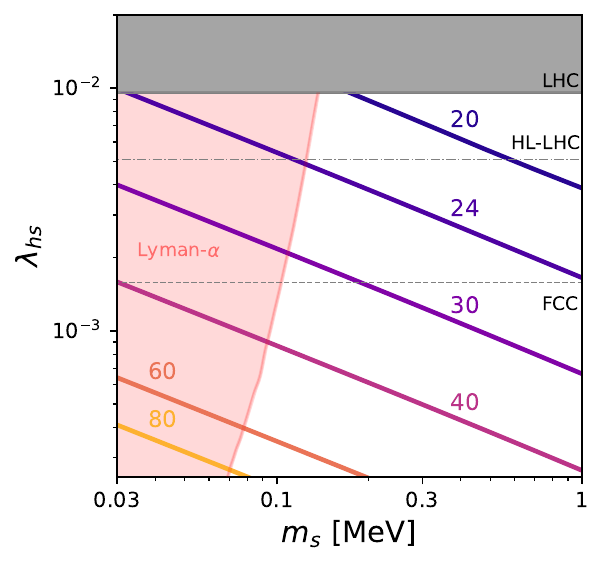}
    \caption{Parameter space of the Higgs portal DM model at low masses.  
    Along the colored lines,    the observed DM abundance is reproduced, for a given $T_R$ (in MeV).
    The shaded areas are excluded by the LHC and Lyman-$\alpha$ constraints, while the dashed lines show the  prospects of
    DM detection via invisible Higgs decay.}
    \label{par}
\end{figure}

   Fig.\,\ref{par} shows the result of our parameter space analysis. We are focussing the low dark matter masses below 1 MeV since for larger masses our scalar DM is cold \cite{Lebedev:2024mbj}. We also take the maximal reheating temperature 
   to be 80 MeV in order to stay safely below the QCD phase transition temperature, close to which our perturbative dilute gas analysis becomes inadequate. 
   The observed DM abundance is reproduced along the lines marked by $T_R$ (in MeV).\footnote{These agree within a factor of 1.4 to 1.9 with our previous results \cite{Lebedev:2024mbj}. The discrepancy  is due to the different 
   treatment of the hadronic states: numerical integration over various hadronic states based on a plot in \cite{Winkler:2018qyg} is now replaced with the chiral perturbation limit which only involves the pions at low $T$. We note that our treatment of the SM plasma as an ideal gas
   also entails some uncertainty, especially at temperatures close to the QCD phase transition \cite{HotQCD:2012fhj}.}

   The pink shaded area is excluded by the Lyman-$\alpha$ constraints on warm DM, which we have analyzed using the CLASS tool \cite{Blas:2011rf,Lesgourgues:2011rh} as described in Appendix A. These become stronger for smaller $T_R$. Indeed, the DM momentum at the production point
   is given by the pion and muon masses, and a lower $T_R$ implies less time for ``red-shifting'' the momentum to small values at the structure formation epoch. This makes dark matter ``warmer''. 
   We observe   that  values of $m_s$ below 70 - 100 keV are excluded. Such a  constraint is stronger than that in  high-$T$ models due to the larger   DM comoving momentum, $q \sim m_{\pi, \mu} / T_R $, and the IR suppression
   of the momentum distribution.

   The grey area is excluded by the LHC search for  invisible Higgs decay, $h\rightarrow SS$. The corresponding decay width is given by 
   \begin{equation}
 \Gamma (h \rightarrow SS)= { \lambda^2_{hs} v^2 \over 32 \pi m_h} \sqrt{1- {4m_s^2 \over m_h^2}} \;.
 \end{equation}
   We use the bound on the invisible decay branching ratio BR$_{\rm inv} \lesssim 10$\% \cite{ATLAS:2023tkt}, which 
   excludes the Higgs portal couplings above $10^{-2}$.
   The dashed lines show the prospects for observing this mode at the HL-LHC and FCC.
   We take the HL-LHC benchmark goal to be BR$_{\rm inv} = 3$\% \cite{RivadeneiraBracho:2022sph}, while that of the FCC
to be BR$_{\rm inv} = 0.3$\% \cite{FCC}.
    
   We thus conclude that warm dark matter in the Higgs portal framework can be probed  via invisible Higgs decay at current and future colliders.

  \subsection{Models with a flat temperature profile before reheating}
  
  Consider now a class of models exhibiting a flat temperature profile before reheating. This corresponds to $n-m=0$ in Eq.\,\ref{system} and  curve (2) in Fig.\,\ref{Tmax}. 
  Such a possibility is realized, for example, when the inflaton decays into the right-handed neutrinos $\nu_R$, which subsequently produce the SM fields: $\phi \rightarrow \nu_R \nu_R \rightarrow {\rm SM}$.

  \begin{figure}[h!]
    \centering
    \includegraphics[width=0.55\textwidth]{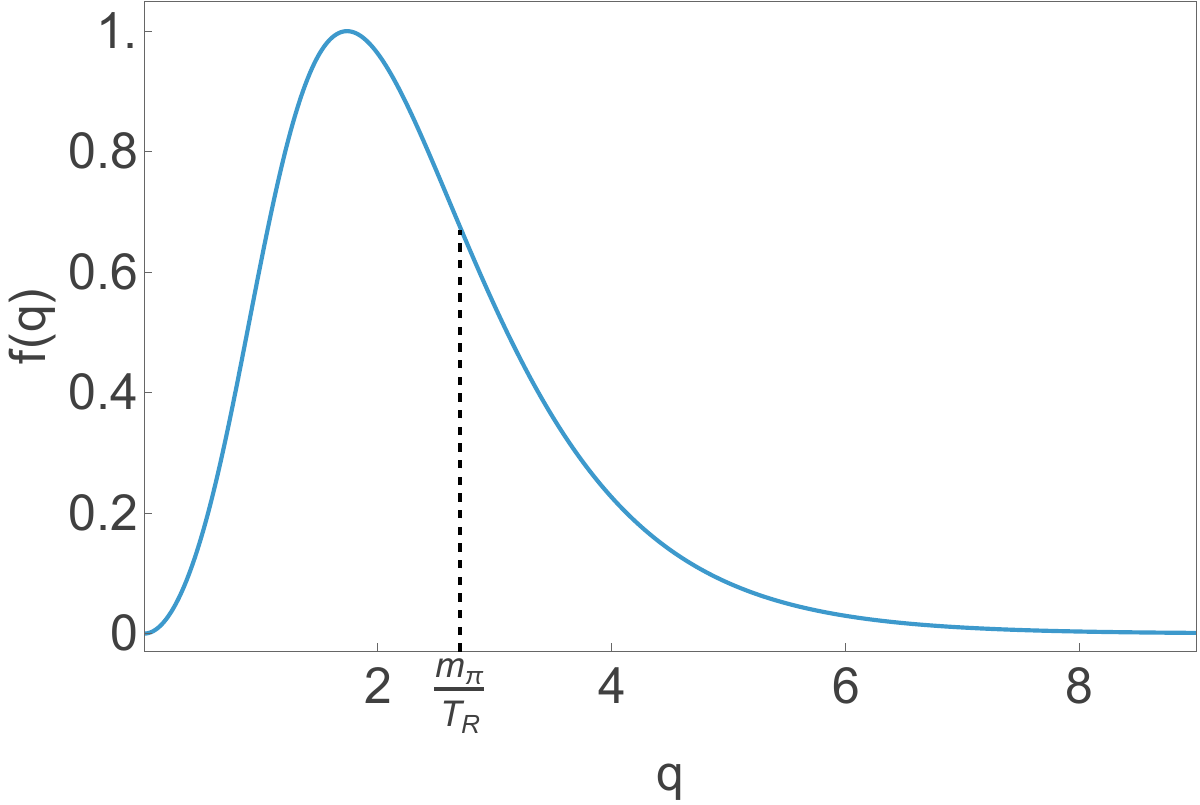}
    \caption{Comoving momentum distribution function of DM produced via pion annihilation, with $T_R=50\;$MeV and assuming a constant $T$ profile before reheating.   
    For convenience, $f(q)$  is normalized to 1 at the maximum.}
    \label{f1}
\end{figure}
  
  Compared to the previous case, the momentum distribution function receives an extra piece $\Delta f$ which is due to the particles produced before the reheating point at $a=a_R$. The full distribution function is given by
  \begin{equation}
  f(q) = f(q)\Big\vert_{T<T_R} + \Delta f (q) \; ,
  \end{equation} 
  where $f(q)\Big\vert_{T<T_R} $ has been computed in the previous subsection and $\Delta f$ is 
    \begin{equation}
\Delta f(p,a_R) ={1\over p\, a_R}\; \int^{a_R}_{0} {da' \over H(a')} \, {{\cal C}(p',a')  } ~~{\rm with}~~ { p}' = {{ p} \, {a_R \over a'}}\;.
\end{equation}
In this integral, the temperature is taken to be constant, $T=T_R$ and 
the comoving momentum is given by
$q=p/T_R$ at $a=a_R$. After that, the comoving momentum  $q=p(T)/T$ remains constant. Before reheating, we may take $H\propto a^{-2}$ assuming radiation-like scaling of the energy density of the Universe,
although this is not a critical assumption and analogous results can readily be obtained for different scaling laws.

The asymptotic behavior of $f(q)$ can be understood analytically.
  Consider the low momentum regime $p \ll T_R$. The collision term can be approximated by ${\cal C}(p') \propto p^{\prime \alpha} \exp (-p'/ T_R -m_\pi^2/ (p' T_R))$
  with some positive $\alpha$. The integral is dominated by the vicinity of the point $a'_*= a_R \;p /m_\pi$ which maximizes the exponential. Around this point, the exponential  has a Gaussian form
  with a small width $\propto p$ such that the integrand  can be approximated by a ``box'' function. This yields $\Delta f \propto p^2$, which also gives the asymptotic behavior of the full function $f$. 
 In the  large $p$ limit, the above integral is suppressed compared to $f(q)\Big\vert_{T<T_R}$ and the asymptotic behavior of $f(q)$ remains the same as in the previous section.
  
  These results are dictated by simple physical considerations. The DM quanta with arbitrarily small momenta $p$  are generated via ``red-shifting'' the initially energetic particles ($E\simeq m_\pi$). For a given $p$, this fixes 
  the production scale factor $a'_*= a_R \;p /m_\pi \ll a_R$. On the other hand, production of highly energetic particles is dominated by late times. Indeed, the temperature remains constant between $a\sim 0$ and $a=a_R$, while particles produced at earlier times are subject to the ``red-shifting''. Hence, the constant $T$ period does not affect the large momentum behavior of the distribution function.

    We thus find the following asymptotic scaling of $f(q)$:
  \begin{eqnarray}
  && q\ll 1~:~~~~~~~~~~ f(q) \propto   q^2   \;,\nonumber \\
  && q\gg m_\pi^2/T_R^2 ~:~~  f(q) \propto  {1\over \sqrt{q} }\; \exp\left(   -q  \right)\;.
  \label{f(q)1}
  \end{eqnarray}
Our numerical result is presented in Fig.\,\ref{f1}. Compared to our previous findings, we observe that the peak of the distribution moves to a lower momentum and that the IR modes
do not get cut off.
  Nevertheless, the above asymptotic form shows that the $\alpha\beta\gamma$-parametrization is still not applicable, strictly speaking.\footnote{We find that the $\alpha\beta\gamma$-parametrization can reproduce well the overall shape of $f(q)$, while
  failing in the asymptotic regions. This difference may be insignificant for applications.} 
  This is natural since the low and high momentum behavior is determined by the pre-reheating and post-reheating epochs, respectively.
  We also find  that the peak of the distribution is correlated with $m_\pi/T_R$
  and, in practice, located not far from $q\sim m_\pi/T_R$. An analogous  statement applies to $f(q)$ generated in the muon annihilation channel.

 \begin{figure}[h!]
    \centering
    \includegraphics[width=0.5\textwidth]{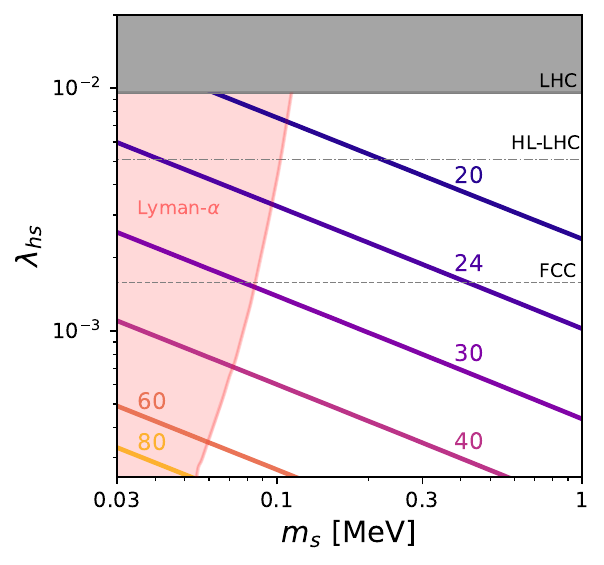}
    \caption{Parameter space of the Higgs portal DM model at low masses, assuming a constant temperature profile before reheating.  
    Along the colored lines,    the observed DM abundance is reproduced, for a given $T_R$ (in MeV).
    The shaded areas are excluded by the LHC and Lyman-$\alpha$ constraints, while the dashed lines show the  prospects of
    DM detection via invisible Higgs decay.}
    \label{par-1}
\end{figure}

   The DM relic abundance now contains two contributions: one from the SM radiation domination epoch ($a>a_R$), as before, and the other from the  pre-reheating era ($a<a_R$).
   The new contribution is 
    \begin{equation}
  \Delta Y = {2\, \Gamma({\rm SM} \rightarrow SS) \over 5 \,H\, s_{\rm SM} }\bigg\vert_{T=T_R}\;,
  \end{equation}
  where $\Gamma = \langle \sigma v_r \rangle \, n_{\rm SM}^{\rm eq\;2}$ is the DM production rate per unit volume. 
  This results in a smaller coupling being sufficient for producing the right relic density.

  Our parameter space analysis is presented in Fig.\,\ref{par-1}.  The relic abundance lines move somewhat to the lower couplings, with the 
  shift getting smaller 
   at higher temperatures  due to  $\Delta Y/ Y_0 \propto 1/T_R$. The Lyman-$\alpha$ constraints relax compared to the previous case, with the lower bound on the DM mass 
   being in the range $50-100\,$keV. 
   Although the DM momentum distribution shifts to the lower values, most DM particles are still energetic, with typical comoving momenta not far from $m_\pi/T_R$.\footnote{The electron annihilation channel becomes important only at $T_R\lesssim 15\,$MeV.} As a result, the 
   structure formation  bound remains strong.

  \section{Conclusion}
  
  We have studied light Higgs portal  dark matter in the context of freeze-in at stronger coupling. 
  In particular, we have focused on the warm dark matter window and the reheating temperature below 100 MeV.
  The main production channel is the muon and pion annihilation mediated by the Higgs. The reaction rate is suppressed
  by the Boltzmann factor at $T < m_{\mu,\pi}$,  the small Yukawa couplings as well as  by the Higgs propagator. This makes the 
  production mechanism to be of the freeze-in type, even though the Higgs-DM coupling can be significant.

  We find that the observed relic density can be produced in the warm dark matter regime for the Higgs portal coupling 
  $\lambda_{hs}$ in the range $10^{-4}- 10^{-2}$, which leads to the invisible Higgs decay within the reach of the LHC and FCC.
    The Lyman-$\alpha$ bound sets a strong lower bound on the warm DM mass, not far from 100 keV. This is due to the 
    fact that  the DM momentum distribution function has a highly non-thermal  shape, with most particles having  
    momenta substantially above the SM temperature.  Generally, we find that the momentum distribution in the low $T$ freeze-in framework
    is not captured by the usual $\alpha\beta\gamma$ parametrization. 
    
    The model is best probed by the invisible Higgs decay into dark matter,  with the decay branching ratio at the percent level. This mode
    can be observed at the LHC and future colliders such as the FCC.
  \\ \ \\
 {\bf Acknowledgements.} We are grateful to Deanna Hooper for help with the CLASS package.

\appendix

\section{Implementation of constraints in CLASS}

To obtain the Lyman-$\alpha$ constraints using CLASS \cite{Blas:2011rf}, two main inputs are needed: the distribution function $f(\bar p)$ in terms of a dimensionless momentum $\bar p$ and a dimensionless normalization scale $ T_{\rm ncdm}$.
The momentum  $\bar p$ is defined by 
  $\bar p = \frac{p}{{T_{\rm ncdm}} T_0}$, where $p$ is the physical momentum today and $T_0$ is the average CMB temperature today. The most convenient normalization scale for freeze-in at stronger coupling is ${T_{\rm ncdm}} =  \frac{m_\pi}{T_R}$.
  The distribution function can be computed  by solving numerically the Boltzmann equation or by using more sophisticated tools like DRAKE \cite{Binder:2021bmg}.

With the above input, CLASS computes the power spectrum corresponding to a given dark matter model. Then, the Lyman-$\alpha$ constraints can be imposed using the area criterion, detailed in \cite{Decant:2021mhj} and \cite{Ballesteros:2020adh}.
To apply this criterion, one first needs to obtain the one-dimensional power spectrum  by integrating the three-dimensional one over the momentum, from $k$ to $k_{\rm lim} \to \infty$. Then, one quantifies the difference  between our non-thermal DM model and the cold DM case through the ratio $r(k)$ of the two spectra. 
  The area is then defined as $A_X= \int_{k_{\rm min}}^{k_{\rm max}} dk \, r(k)$,
where $k_{\rm min} = 0.5 \, h/{\rm Mpc}$ and $k_{\rm max} = 20 \, h/{\rm Mpc}$ \cite{Decant:2021mhj}. In the CDM case,  $A_{\rm CDM} = k_{\rm max} - k_{\rm min}$. 
At the next step, one obtains the area estimator   $\delta A_X= \frac{A_{\rm CDM} - A_X}{A_{\rm CDM}}$.
This estimator must be first computed for a benchmark WDM model with the Fermi-Dirac distribution function and a DM mass of $5.3 \,\rm keV$. Then, the same estimator is computed for our DM model. In order for our model to be consistent with the Lyman-$\alpha$ observations, the area estimator must be below that of the benchmark model.
The choice of $k_{\rm lim}$   can have an impact on  the area estimator as shown in \cite{DEramo:2020gpr}. However, the dependence on $k_{\rm lim}$ is  the same for any relevant model. Thus, as long as we choose the same value of $k_{\rm lim}$ for all the calculations, we obtain consistent results. 

Following this procedure, we find the benchmark model value
\begin{equation}
     \delta A_X^{\rm lim} =  0.205 , \quad m_{\rm WDM} = 5.3 \, {\rm keV}, \quad k_{\rm lim} = 100 \,h/{\rm Mpc}\,.
     \label{eq:boundA}
\end{equation}
To identify the region   allowed by the Lyman-$\alpha$ observations, we apply the following procedure. 
\begin{enumerate}
    \item Choose $T_R$ and compute $f(\bar p)$. ${ T_{\rm ncdm}}$ and $f(\bar p)$ are input in  CLASS.
    \item Choose  $m_{\rm DM}$  and input it in CLASS.
    \item Compute the area estimator and check if it is bellow the bound  in Eq.~\eqref{eq:boundA}.
    \item Vary $m_{\rm DM}$ until  the area estimator  bound is saturated.  Lower values of  $m_{\rm DM}$ are excluded.
    \item Repeat for a different $T_R$.  A pair ($m_{\rm DM}$, $T_R$) corresponds to a point in the parameter space ($m_{\rm DM}$, $\lambda_{hs}$) through the relic density condition. It lies on the boundary of the allowed region.
\end{enumerate}

\end{document}